\documentstyle[12pt]{article}

\begin{document}
\newcommand{\od}{\stackrel{\cdot}}
\newcommand{\td}{\stackrel{\cdot\cdot}}
\newcommand{\fd}{\stackrel{\cdot\cdot\cdot}}
\topmargin 0pt
\oddsidemargin 5mm
\begin{titlepage}
\setcounter{page}{0}
\rightline{Preprint BUTP-98/14}
\vspace{2cm}
\begin{center}
{\Large Edge Excitations of an Incompressible Fermionic Liquid 
in a Disorder Magnetic Field}
\vspace{1cm}
{\large} \\
{\large  A. Sedrakyan\footnote{e-mail: {\sl sedrak@lx2.yerphi.am;
 sedrak@nbivax.nbi.dk}}}

%\vspace{1cm}
%{\large t}

{ITP, University of Bern, Sidlerstrasse 5, CH-3012, Bern
\footnote{Permanent address: Yerevan Physics Institute,
Br.Alikhanian st.2, Yerevan 36, Armenia}}
\end{center}
\vspace{5mm}
\centerline{{\bf{Abstract}}}

 The model of lattice fermions in 2+1 dimensional space is
formulated, the critical states of which are lying in the basis
of such physical problems, as 3D Ising Model(3DIM) and the edge
excitations in the Hall effect.
The action for this exitations coincides with the action
of so called sign-factor model in 3DIM at one values
of its parameters, and represent a model for the edge 
excitations, which are responsible for the plato transitions
in the Hall effect, at other values.
The model can be formulated also as a loop gas models in 2D,
but unlikely the $O(n)$ models, where the loop fugacity is
real, here we have directed (clochwise and conterclochwise) 
loops and phase factors $e^{\pm 2\pi {p \over q} i}$ 
for them. The line of phase transitions in the parametric
space will be found and corresponding continuum limits of this 
models will be constructed.
It appears, that besides the ordinary critical line, which 
separates the dense and diluted phases of the models(like
in ordinary $O(n)$ models),  there is a line, which corresponds
to the full covering of the space by curves. The $N=2$
twisted superconformal models with $SU(2)/U(1)$ coset model
coupling constant $k={q \over p}-2$ describes this states. 

\end{titlepage}
\newpage
\renewcommand{\thefootnote}{\arabic{footnote}}
\setcounter{footnote}{0}

\section{Introduction} 
\indent

It is well known that in most of important problems of
modern physics, such as quark confinement in Quantum
Chromodynamics (QCD) or high temperature superconductivity,
the understanding and description of ground state is crucial
for the construction of the theory of corresponding phenomena.
For a systems in the strong coupling regime the ground state
is far from trivial vacuum, as we have in the perturbative 
theories, being essentially structured. Hence, the important 
step in developing such theories should be the understanding
of the ground state.

The aim of this article is the formulation of the lattice model
in 2+1 dimensional space and investigation of its critical
states, which,as it seems to us, is lying in the basis of such 
physical
problems as $3D$- Ising model ($3DIM$) \cite{PD,S,KS} and
edge excitations, responsible for the plato transitions in
the Hall effect \cite{PG,H,WEN}.

In the presented model we have used the essential ingredients 
of the model for so-called sign-factor, formulated in the string
representation if the $3DIM$ \cite{KS}, namely, the
chiral fermions, which are hopping along the arrows on the 
Manhattan lattice ($ML$) (Fig.1) in the presence of some
external $SU(2)$ gauge field.

The external gauge field in \cite{KS} was induced by the
random surface in the $3D$ regular lattice and consists
on background  fluxes per plaquette, which takes values in the $Z_2$
center of the $SU(2)$ group.

Now, simplifying the situation, we will consider the
tight-binding model for electrons, which are hopping along
the arrows of $ML$ in the external $U(1)$ field, which formed
by $\pm 2 \pi p/q$ fluxes per plaquette, arranged in a
chess-like order. The model describes the motion of
electrons in a regular lattice of magnetic vortices with
the staggered directions of fluxes, which are placed in 
the middle points of plaquettes.

Being interested in a low lying ( $E \approx 0$) critical
excitations of the model we are changing the creation and
annihilation operators of electrons in the proposed
Hamiltonian by Grassmann valued fields in two dimensional 
space and obtain the action for them. By taking functional
integral over this Grassmann variables we will find
an alternative 
 formulation for the partition function as loop gas model,
expressed as a statistical sum of random non self-intersecting
curves (due to $ML$ structure), which appeared with the usual
Boltzmann weights as $e^{m_0 L}$ and fugacities equal to phase 
factor $\eta =- e^{\pm 2\pi {p \over q}i}$.  One
should not mix this models with the usual loop-gas
(O(n) or Q-Potts) models \cite{DD1,DK}, where the fugacities
are real and contours are not oriented.

In this article we will consider the limit ${m \over t} 
\rightarrow 0$, where $m$ is the chemical potential and $t$ is 
the hopping parameters of electrons, when the contours
in loop gas formulation of the model fully cover the $2d$
 space (or random surface in general case). The limits
$m \rightarrow 0$ and $t \rightarrow \infty$ are equivalent
after corresponding rescaling of the partition function
and we will refer it as fully packed phase. It will be shown 
that in this limit the model of low lying states is equivalent 
to twisted $N=2$ 
superconformal field theory \cite{MSS,I},
 the characteristic $SU(2)/U(1)$
coset model coupling constant of which is $k= {q \over p}-2$. 
As it is well
known \cite{EY,VV} this twisted models are topological
\cite{W,OSV}, so we have a statistical physics models corresponding to 
them. 

For  sign-factor of the $3DIM$ \cite{KS} it is necessary to consider
fermions in fully packed phase and in the $\Phi=\pi$-flux
background, which means that fugacities of the electronic paths
are equal to one. But for the full construction of sign factor it is
necessary to add to  the action a pair of 
vertex operators, which corresponds to the creation of additional 
impurity fluxes $\Phi = \pi$ at some points. This fluxes
are induced by the Whitney singularities \cite{WH} of the immersed
with the defects $2D$ manifolds into the $3D$ Euclidean
space, which appeared in sign-factors construction of the 
$3DIM$ \cite{KS}. Here we will not consider  impurity
fluxes, trying first to construct the continuum limit of the model
without them, but proceeding in this way we hope to have a
topological description of the sign-factor in terms of the 
geometry of surfaces.

Another interesting application of our model is Hall effect,
which corresponds to another point of parametric space $(m, t)$
The main characteristics of the Hall effect is the fact 
that  the
states in the bulk are forming incompressible liquid of localized
electrons \cite{L},  while there is delocalized  
states near
the edges of the sample, which at the appropriate density
of electrons (or at some external magnetic fields) becomes
occupied as a Fermi level and, hence, is responsible for the
longitudinal conductivity at the plato transition points.
It  appears, that our model demonstrates similar future.
Moreover, by constructing the transfer matrix for the 
low lying excitations in our model, one can recognize the
transfer matrix, proposed by Chalker and Coddington
in \cite{CC} for scattering of the
edge excitations. It will be shown \cite{AS}, this long range
excitations can be described as a (0,1)-spin fermionic
system with the central charge $c=-2$.

In the Section 2 we present the description of the model 
and calculate the spectrum of excitations.
 
In the Section 3 we construct the transfer matrix and calculate 
the spectrum of the model on $ML$ in the absence of the 
background fluxes. In the appropriate scaling limit
of the parameters of the model we will have a massless
(0,1)-spin fermionic system with the central charge
$c=-2$.

Section 4 intended for the analyze of critical states
and on representation of background $U(1)$ fluxes via
vertex operators of two fluctuating scalar fields with
opposite statistics. The construction of the continuum 
limit of the
whole model for the low lying excitations will be completed
in the Section 5.

It is worth to mention now, that proposed in 
\cite{KS} and here Hamiltonian 
is non-Hermitian. Recently  non-Hermitian Hamiltonians
have attracted much attention in respect with vortex pinning
problem in superconductors \cite{HB}, directed quantum chaos
\cite{EF} and in study of sliding of charge density waves
in disorder systems \cite{CBF}.
%%%%%--------------------------------------------------
%%%%%------------------------------------------------
%%%%%%------------------------------------------------

\section{Description of the Model}
\indent
 The Manhattan lattice ($ML$) is the lattice, where there are
continuous arrows on the links with the opposite directions
on the neighbor parallel lines (Fig.1). The arrows form a set of 
vectors
$\vec{\mu}_{ij}\in {\cal S}$. $ML$ originally was defined by Kasteleyn
\cite{K} in connection with the problem of single Hamiltonian
 walk (HW).

\begin{center}
\small{
\setlength{\unitlength}{6mm}
\begin{picture}(15,13)
\multiput(0,2)(3,0){5}{\vector(1,0){3}}
\multiput(3,5)(3,0){5}{\vector(-1,0){3}}
\multiput(0,8)(3,0){5}{\vector(1,0){3}}
\multiput(3,11)(3,0){5}{\vector(-1,0){3}}
\multiput(1.5,0)(0,3){4}{\vector(0,1){3}}
\multiput(4.5,3)(0,3){4}{\vector(0,-1){3}}
\multiput(7.5,0)(0,3){4}{\vector(0,1){3}}
\multiput(10.5,3)(0,3){4}{\vector(0,-1){3}}
\multiput(13.5,0)(0,3){4}{\vector(0,1){3}} 
\put(3,3.5){\shortstack{\circle{0.2}}}
\put(3,6.5){\shortstack{\circle*{0.2}}}
\put(6,3.5){\shortstack{\circle*{0.2}}}
\put(6,6.5){\shortstack{\circle{0.2}}}
\put(9,3.5){\shortstack{\circle{0.2}}}
\put(9,6.5){\shortstack{\circle*{0.2}}}
\put(12,3.5){\shortstack{\circle*{0.2}}}
\put(12,6.5){\shortstack{\circle{0.2}}}
\put(3,9.5){\shortstack{\circle{0.2}}}
\put(6,9.5){\shortstack{\circle*{0.2}}}
\put(9,9.5){\shortstack{\circle{0.2}}}
\put(12,9.5){\shortstack{\circle*{0.2}}}
\put(2.5,9){\shortstack{$B_1$}}
\put(5.5,9){\shortstack{$A_2$}}
\put(8.5,9){\shortstack{$B_1$}}
\put(11.5,9){\shortstack{$A_2$}}
\put(3.8,8.3){\shortstack{2}}
\put(6.8,8.3){\shortstack{3}}
\put(9.8,8.3){\shortstack{2}}
\put(12.8,8.3){\shortstack{3}}
\put(3.8,5.3){\shortstack{1}}
\put(6.8,5.3){\shortstack{4}}
\put(9.8,5.3){\shortstack{1}}
\put(12.8,5.3){\shortstack{4}}
\put(3.8,2.3){\shortstack{2}}
\put(6.8,2.3){\shortstack{3}}
\put(9.8,2.3){\shortstack{2}}
\put(12.8,2.3){\shortstack{3}}
\put(2.5,3){\shortstack{$B_1$}}
\put(5.5,3){\shortstack{$A_2$}}
\put(8.5,3){\shortstack{$B_1$}}
\put(11.5,3){\shortstack{$A_2$}}
\put(2.5,6){\shortstack{$A_1$}}
\put(5.5,6){\shortstack{$B_2$}}
\put(8.5,6){\shortstack{$A_1$}}
\put(11.5,6){\shortstack{$B_2$}}
\put(6.5,-2.5){\shortstack{Fig.1.}}
\put(3.8,-1){\shortstack{2j-1}}
\put(1,-1){\shortstack{2j-2}}
\put(7,-1){\shortstack{2j}}
\put(10,-1){\shortstack{2j+1}}
\put(-2,2){\shortstack{m}}
\put(-2,5){\shortstack{m+1}}
\end{picture}}
\end{center}

\vspace{15mm}

 The multiple HW problem was considered and solved by Duplantie 
and David
\cite{DD}. Here we will consider flat $ML$, though the random
scarface with the $ML$ structure, which is necessary to have for
$3DIM$, also can be defined easily \cite{KS,DK}.

The plaquettes of $ML$ are divided into four groups, $A_a$ and $B_a$
 (a=1,2),
destinated in the chess like order. The A-plaquettes differ from
B-plaquettes by the fact, that arrows are circulating around them,
while there is no circulation for B-plaquettes. $A_1(A_2)$ has 
clockwise(counterclockwise) circulation, while $B_1$ differs from the
$B_2$ by rotation on $\pi/4$ (see fig.1).  We will organize
the walk of the fermions along the arrows with hopping parameters,
which are invariant on double lattice spacings translations, 
simultaneously being
in the disordered $U(1)$ (magnetic) field. The
magnetic field is defined as follows.  There is
$U(1)$ flux $\Phi=2\pi p/q$ in the A-plaquettes 
and $\Phi=-2\pi~p/q$, in the B-plaquettes 
($p$ and $q$ are mutually prime integer
numbers). On the random $ML$ only the $A$-type plaquettes
can be deformed into $n$-angles, while $B$-type plaquettes will 
stay as quadrangles.

We will consider one fermionic degree of freedom per lattice
site. Because of $ML$ structure the translational invariance
occur only for double lattice spacing translations in both
directions, the corresponding Brillouin zone of fermions will
be reduced and we will have four bands. 
Let $C^{i+}_{\vec{n}}, C^i_{\vec{n}},\,(i=1,2,3,4)$ are 
corresponding creation and annihilation operators at the 
lattice sites  $\vec{n}=(n,m)$. For example (see Fig.1) we can
mark operators as follows
\begin{eqnarray}
C^1_{\vec{n}}\quad&{\rm at\,\, the}
&\quad\vec{n}=({\rm odd,\, odd})\nonumber\\
C^2_{\vec{n}}\quad&{\rm at\,\, the}
&\quad\vec{n}=({\rm odd,\, even})\nonumber\\
C^3_{\vec{n}}\quad&{\rm at\,\, the}
&\quad\vec{n}=({\rm even,\, even})\nonumber\\
C^4_{\vec{n}}\quad&{\rm at\,\, the}
&\quad\vec{n}=({\rm even,\, odd})\nonumber
\end{eqnarray}

The Hamiltonian of the model we would like to define is
\begin{eqnarray}
\label{ACT}
{\cal H}(C_{\vec{n}}^i;\Phi)&=& \sum_{\vec{n},\vec{\mu }\atop{\rm
on}\,\,ML}\left(t_{14} C^{1+}_{\vec{n}}
U_{\vec{n},\,\vec{\mu}}C^4_{\vec{n}+\vec{\mu}} +t_{43}
C^{4+}_{\vec{n}}U_{\vec{n},\,\vec{\mu}}C^{3}_{\vec{n}+\vec{\mu}}+
t_{32}C^{3+}_{\vec{n}}U_{\vec{n},\,\vec{\mu}}
C^{2}_{\vec{n}+\vec{\mu}}+\right.
\nonumber\\
&+&\left.t_{21} C^{2+}_{\vec{n}}U_{\vec{n},\,\vec{\mu}}C^{1}_{\vec{n}+
\vec{\mu}}+ t_{41} C^{4+}_{\vec{n}}
U_{\vec{n},\,\vec{\mu}}C^1_{\vec{n}+\vec{\mu}} +t_{34}
C^{3+}_{\vec{n}}U_{\vec{n},\,\vec{\mu}}C^{4}_{\vec{n}+\vec{\mu}} +\right.
\nonumber\\
&+&\left.t_{23}C^{2+}_{\vec{n}}U_{\vec{n},\,\vec{\mu}}
C^{3}_{\vec{n}+\vec{\mu}}
+ t_{12} C^{1+}_{\vec{n}}U_{\vec{n},\,\vec{\mu}}C^{2}_{\vec{n}+
\vec{\mu}} \right) + m \sum_{{\rm sites} \atop{i=1,..4}} C^{i+}_{\vec{n}} 
C^i_{\vec{n}}
\end{eqnarray}
where $\vec{\mu}$ are the unit vectors on $ML$ along of directions of the
arrows, $m$ is the chemical potential and relations
\begin{eqnarray}
\label{UD}
&&U_{\vec{n},\vec{\mu}}\in U(1),\quad \eta_A=\eta=\prod_A
UUUU=e^{i\Phi}=e^{2\pi ip/q},\nonumber\\
&&\eta_B={\eta}^{-1}=\prod_B UUUU =e^{-i \Phi}
\end{eqnarray}
defines the external $U(1)$ magnetic field background. 
The fermion hopping amplitudes
$t_{ij}$ can be regarded as
 metric elements on the
lattice  and may be represented as 
$\rho e^{\Gamma_{\vec{n},\,\vec{\mu}}}$, where $\rho$ is the conformal
factor and  $\ln \rho+\Gamma_{\vec{n},\,\vec{\mu}}$ can be regarded
as gravitational connection.
 It is important to mention the absence of 
the complex conjugate terms in the expression (\ref{ACT})
and therefore the Hamiltonian is non-Hermitian.

So we have the Hamiltonian for fermions, which are living on
Manhattan lattice in the  disordered external
magnetic and gravitational backgrounds.

One can write the partition function for the excitations 
of energy $E$ of the model as
\begin{equation}
\label{STSN}
Z(E)=\int \prod_{i=1}^{4} \prod_{\vec{n}}\,dC_{\vec{n}}^i\,
d \bar{C}_{\vec{n}}^i\,e^{-{\cal H}(C_{\vec{n}}^i;\Phi) + 
E \sum_{{\rm sites} \atop{i=1,..4}} C^{i+}_{\vec{n}} 
C^i_{\vec{n}}}
\end{equation}
where $C_{\vec{n}}^i$ and $\bar{C}_{\vec{n}}^i$ are now independent
Grassmann fields, corresponding to $C_{\vec{n}}^i$ and 
$C_{\vec{n}}^{i+}$.

In the Bloch wave basis and after some simple gauge fixing the
Hamiltonian (\ref{ACT}) can be written as
\begin{eqnarray}
\label{BL}
{\cal H}(C_{\vec{n}}^i;\Phi) = \sum_{\vec{k}}C^{i+}_{\vec{k}} 
{\cal H}(\vec{k})_{ij}C^{j}_{\vec{k}},
&& k_x= {2\pi \over L} l, l=0,....{L \over 2}-1,\\ 
&&k_y= {2\pi \over N} n, n=0,....{N \over 2}-1,\nonumber 
\end{eqnarray}
where ${\cal H}(\vec{k})_{ij}$ are the elements of the following 
matrix 
\begin{eqnarray}
\label{KIKO}
{\cal H}(\vec{k})=\left(\begin{array}{llll} m&\omega 
t_{12} e^{-ik_y}&
0&\omega^{-1} t_{14} e^{-ik_x}\\
\omega^{-1} t_{21} e^{-ik_y}& m& \omega t_{23} e^{ik_x}& 0\\
0 &\omega^{-1} t_{32} e^{ik_x} & m &\omega t_{34} e^{ik_y}\\
\omega t_{41} e^{-ik_x} & 0,&\omega^{-1} t_{43} e^{ik_y} & m 
\end{array}\right),
\end{eqnarray}
and $\omega=e^{\frac{i \pi}{2}\frac{p}{q}}, \omega^4=\eta$.

It's not hard now to find the spectrum of the model (1) by 
 diagonalizing  $4\times 4$ matrix of ${\cal H}(\vec{k})$.
The answer is obviously invariant under double lattice  space{\em \/}
periodic $U(1)$ transformations and can be written as
\begin{eqnarray}
\label{ZX}
E=m&{\pm}&\left\{\frac{1}{2}\left(t_{23} t_{32} e^{2i k_x}+ 
t_{14} t_{41} e^{-2i k_x}+t_{43} t_{34} e^{2i k_y}+t_{12} t_{21} 
e^{-2i k_y}\right) \pm\right.
\nonumber\\
&\pm&\left[\frac{1}{4}\left(t_{23} t_{32} e^{2i k_x}+ 
t_{14} t_{41} e^{-2i k_x}+t_{43} t_{34} e^{2i k_y}+t_{12} t_{21} 
e^{-2i k_y} 
\right)^2 - \right.
\nonumber\\
&-&\left. \left. (t_{14} t_{32} -
\eta t_{12} t_{34})(t_{23} t_{41}-\eta^{-1}  
t_{43} t_{21})\right]^{1/2}\right\}^{1/2}
\end{eqnarray}

One can see, that when $m$=0 we have a $E \rightarrow -E$ symmetry
in the spectrum of the model, which is consequence of 
anticommutativity of the Hamiltonian ${\cal H}(\vec{k})$ with
the chirality operator $\Gamma$
\begin{equation}
\label{QWE}
\left\{ {\cal H}(\vec{k}), \Gamma \right\}= 0, \quad \quad
\Gamma=\left(\begin{array}{ll} 0& I\\
I& 0 
\end{array} \right).
\end{equation}
 $I$ is a $2 \times 2$ unit matrix.

Let as now to investigate the critical $E\approx 0$ states of the system
and consider homogeneous $t_{ij}=t$ case for simplicity. It is
easy to find from (\ref{ZX}), that $E=0$ for
\begin{equation}
\label{E11}
\cos 2 k_x +\cos 2k_y ={m^2 \over {2 t^2}}+ 2 (\sin {\Phi \over 2})^2 
{t^2 \over m^2},
\end{equation}
and the only solutions of this equation in the limit 
${m \over t} \rightarrow 0$ are  imaginary $k_x$ (or $k_y$)
with $|k| \rightarrow \infty$. In general, the real solutions
appears by appropriate choose of hopping parameters $t_{ij}$
both, for finite and infinite limits,  only for $\Phi =0$.  
This means that
wave functions of $E=0$ states exponentially grows to the boundary
of the sample, localizing fermions there. One can call this states
as edge states. What we will do further in this article, is just
to prove, that in a case of appropriate choose of hopping parameters,
this edge excitations are massless and find the 
corresponding field theory for them.

The partition function for this edge excitations is simply
$Z(0)$ in (\ref{STSN}), which means that the original
Hamiltonian (\ref{ACT}) in 2+1 space,  written in  
Grassmann fields (instead of creation-annihilation operators),
can be considered as the action for them in $2D$.
\begin{equation}
\label{E12}
A(C_{\vec{n}}^i;\Phi)={\cal H}(C_{\vec{n}}^i;\Phi)
\end{equation}
Without loss of generality we can re-scale Grassmann fields in the
action $A(C_{\vec{n}}^i;\Phi)$  in order to fix  $m=1$. This will require
an appropriate rescaling of the partition function.

%---------------------
Because of Grassmann properties of the fields $C_{\vec{n}}^i$, $Z(0)$ 
is equal
to the sum of products of Wilson loop integrals from $U(1)$ and 
gravitational
fields over all possible oriented
contours, which covers the lattice. The $U(1)$ loop integrals represents
the magnetic fluxes through the contours, which is always equal to $
\Phi=\pm 2\pi ip/q$, due to the disordered character and the fact,
that in ML all closed curves contain odd number of plaquettes.

Therefore
\begin{equation}
\label{STS1}
Z(0)=\int \prod_{i=1}^{4} \prod_{\vec{n}}\,dC_{\vec{n}}^i\,d 
\bar{C}_{\vec{n}}^i\,e^{-A(C_{\vec{n}}^i;\Phi)}
=\sum_{{\rm all\,\,possible\atop coverings}} (-\eta)^{N_1} (-\eta)^{-N_2}
\prod_{\sigma,\tau}\Phi(\gamma_\sigma)\Phi(\bar{\gamma}_\tau),
\end{equation}
where $N_1(N_2)$ is the number of the closed contours $\gamma_\sigma
(\bar{\gamma}_\tau),
(\sigma,\tau=1....N_1(N_2))$ with clockwise
(anti-clockwise) orientations and 
\begin{equation}
\label{STS2}
\Phi(\gamma)=\prod_{\vec{n}\in \gamma}\rho e^{\Gamma_{\vec{n},\,\vec{\mu}}}.
\end{equation}
The sum in (\ref{STS1}) is taken over all possible coverings.
We formally will not make
difference between clockwise and counterclockwise notations for loops
in the future if its not necessary.

In the homogeneous case, when $t_{ij}=t$, one can find a similarity
of (\ref{STS1}) with the critical $Q$-state Potts or the $O(n)$ loop
gas models, when the loop weight factor(the fugacity)
\begin{equation}
\label{FUG}
-\eta=\sqrt{Q}=n,
\end{equation}
(see \cite{NN,DD1,DK}). The essential difference of our model from 
the mentioned ones is the fact, that instead of real loop weight factor 
we have a phase factor $-\eta$, defined by (\ref{UD}).

After corresponding rescaling of $Z(0)$ in the limit ${m \over t}
\rightarrow 0$ by $t^{-L N}$ ($L N$ is the number of $ML$ lattice sites)
only the dense coverings of the lattice by loops will survive (the mass 
term will not contribute) and all loops will appear only with fugacities
$-\eta^{\pm 1}$. The $p/q=1/2$ case is distinguished by its
connection with the sign factor of $3DIM$ \cite{KS}. For this
case we have $-\eta=1$. The generalization of the model for
random surfaces with ML structure on them is straightforward and we 
will discuss it
later. The sign of the surfaces, appeared here, are essentially connected
with the singularities at the endpoints of selfintersection lines of the
immersed surface \cite{KS}. They can be regarded as $U(1)$ flux
impurities, which ensure phase factor (-1) for the each contour
rounding them.

What  immediately follows from (\ref{E11}) is
 that at the
\begin{equation}
\label{XC}
({m \over t})^{2}=2\pm\sqrt{4-(2\sin\Phi/2)^2}=2\pm\sqrt{2+2\cos\Phi}
\end{equation}
we have critical behavior of the free energy of 
$2D$ theory and long range excitations in the spectrum.
Formula (\ref{XC}) very much reminds the well known
formula for $2D$ $O(n)$-loop gas models \cite{NN,DD1,DK} 
phase transition condition \cite{NN}, if
$$
n=-2\cos\Phi
$$

What we are going to demonstrate here is that the model has continuum 
limit (transition line) for all p/q at $t\rightarrow \infty$ too.

%%%%%%%%%%%%%%-----------------------------------------------------------
%%%%%%%%%%%%%%----------------------------------------------------------
%%%%%%%%%%%%%%----------------------------------------------------------
\section{The Spectrum of the Model in the absence of External 
Magnetic Field}
\indent

As we can see from the expression (\ref{E11}), in the limit of
$t \rightarrow \infty$, which needs for a fully packed phase,
there is obvious massless excitations in the limit
$\Phi \rightarrow 0$. This fact determines the way to proceed
further, namely we will consider fermionic fields in the
empty ($\Phi \rightarrow 0$) background and construct his continuum 
limit.
The external $U(1)$ fields and background fluxes will be
included into the model via correlators of two scalar fields,
$\varphi_1$ and $\varphi_2$, which interacts with fermions.

In this section we will analyze the model in case of $\Phi \rightarrow0$.
It is appeared that  the best approach of handling the problem
is the Transfer Matrix approach. Following \cite{FB} let's find quantum 
Hamiltonian on a chain,
such that his coherent-state path-integral along imaginary time
produces action, which coincides with $A(\psi^i;0)$
\begin{equation}
\label{ZET}
Z(\Phi=0) = Tr T^{N/2}=\int\prod_{j,m} d\bar{\psi}_{j,m}d\psi_{j,m} 
e^{-A(\psi^i;0)}
\end{equation} 

From the Fig.1 we see, that due to geometric properties of ML, the
translational invariance of our model is broken for one-lattice spacing
translations and present only for two-lattice spacing translations.
That means the Transfer Matrix we need is the product of two different 
Transfer Matrices, defined on two, $B_1$ and $B_2$, horizontal chains
(choose vertical direction as imaginary time).
\begin{equation}
\label{TT}
T=T_{B_1} T_{B_2} =:e^{-H_{B_1}}::e^{-H_{B_2}}:,
\end{equation}
where :: means some ordering of operators and will be defined below.
The Transfer Matrix will act on a Hilbert space of fermionic states on
a 1D chain.

The geometry of ML shows (see Fig.1) that one can define $T_{B_1}$ on a
chain of $B_1$ plaquettes (and correspondingly $T_{B_2}$- on a $B_2$)
and that is enough to restore the action (\ref{ACT}). The structure
of Transfer Matrices is easy to fix by shrinking vertical lines
(time direction) to points on a chain. Lets define $H_{B_1}$ and
$H_{B_2}$ as follows
\begin{eqnarray}
\label{BB}
-H_{B_1}&=&
\sum_{j}\left(t_{23}c^+_{2j+1}c_{2j}-t_{41}c^+_{2j}c_{2j+1}+M_1c^+_{2j}c_{2j}
+M_{1}^{'}c^+_{2j+1}c_{2j+1}\right), 
\nonumber\\
-H_{B_2}&=&
\sum_{j}\left(-t_{32}c^+_{2j}c_{2j-1}+t_{14}c^+_{2j-1}c_{2j}+
M_2 c^+_{2j}c_{2j}+M_{2}^{'}c^+_{2j+1}c_{2j+1}\right).
\end{eqnarray}
Here $c^+_i$ and $c_i$ are usual fermion creation-annihilation operators,
defined on a vacuum $|0>$. The hopping parameter $t_{ij}$ are the same
as before, while 
$M_{1,2}$ and $M_{1,2}^{'}$ have to be specified later.

Let's define coherent states on the sites of chain as follows:
\begin{eqnarray}
\label{CS1}
|\psi_{2j}\rangle = e^{\psi_{2j}c^+_{2j}}|0\rangle ,\nonumber\\
\langle\bar{\psi}_{2j}| = \langle0| e^{c_{2j}\bar{\psi}_{2j}}
\end{eqnarray}
for the even sites and
\begin{eqnarray}
\label{CS2}
|\bar{\psi}_{2j+1}\rangle = (c^+_{2j+1}-\bar{\psi}_{2j+1})
|0\rangle ,\nonumber\\
\langle\psi_{2j+1}| = \langle 0|(c_{2j+1}+\psi_{2j+1})
\end{eqnarray}
for the odd sites. The Grassmann variables $\bar{\psi}_i ,\psi_i$ 
anti-commutes with operators $c^+_i , c_i$.

This states has following properties
\begin{eqnarray}
\label{PROP}
c_{2j}|\psi_{2j}\rangle =- \psi_{2j}|\psi_{2j}\rangle,\nonumber\\
\langle\bar{\psi}_{2j}|c^+_{2j} = 
-\langle\bar{\psi}_{2j}|\bar{\psi}_{2j},\nonumber\\
c^+_{2j+1}|\bar{\psi}_{2j+1}\rangle = \bar{\psi}_{2j+1}|\bar{\psi}_{2j+1}
\rangle,\\
\langle\psi_{2j+1}|c_{2j+1} =- \langle\psi_{2j+1}|\psi_{2j+1},\nonumber
\end{eqnarray}
with the overlap of two different type of coherent states 
\begin{eqnarray}
\label{TR}
\langle\bar{\psi}_{2j}|\psi_{2j}\rangle =
e^{\bar{\psi}_{2j}\psi_{2j}},\nonumber\\
\langle\psi_{2j+1}|\bar{\psi}_{2j+1}\rangle =
e^{\bar{\psi}_{2j+1}\psi_{2j+1}},
\end{eqnarray}
and the completeness relation of the states as 
\begin{eqnarray}
\label{CR}
\int d\bar{\psi}_{2j}d\psi_{2j}|\psi_{2j}
\rangle \langle\bar{\psi}_{2j}| e^{\psi_{2j}\bar{\psi}_{2j}}&=& 1,
\nonumber\\
\int d\bar{\psi}_{2j+1}d\psi_{2j+1}|\bar{\psi}_{2j+1}\rangle 
\langle\psi_{2j+1}| e^{\psi_{2j+1}\bar{\psi}_{2j+1}}&=& 1.
\end{eqnarray}

The state of the chain at the time $m$ is product of the states at sites
\begin{eqnarray}
\label{PSI}
|\psi(m)\rangle&=&\prod_{j=1}^{L/2}|\psi_{2j}(m)\rangle
|\bar{\psi}_{2j+1}(m)\rangle,\nonumber\\
\langle\psi(m)|&=&\prod_{j=1}^{L/2}\langle\bar{\psi}_{2j}(m)|\langle
\psi_{2j+1}(m)|,
\end{eqnarray}
where $L$ is the length of the chain.

Now the calculation of partition function with Transfer Matrix $T$ by use
of formulas (\ref{TT}-\ref{PSI}) is easy to perform. If we will choose
 usual normal ordering for the fermions at  even points of the chain
and the hole operator ordering at odd points of the chain in the
expression (\ref{TT}) for $T$, then the   
 matrix elements between coherent states, defined
in (\ref{CS1}-\ref{CS2}), can be obtained simply by changing operators
$c_i,c^+_i$ by their eigenvalues $\psi_i,\bar{\psi}_i$. 
This type of ordering means that 
in ground state half of states are filled by fermions.
Then for the partition function we will have 
\begin{eqnarray}
\label{AR}
Z_0&=&Tr T^{N/2} = Tr(T_{B_1}T_{B_2})^{N/2} = \nonumber\\
&=&\int\prod_{j,m} d\bar{\psi}_{j,m}d\psi_{j,m} e^{-S(\bar{\psi},\psi)}
\end{eqnarray}
where
\begin{eqnarray}
\label{SR}
S(\bar{\psi},\psi)&=&\sum_{j,m}\left[t_{23} \bar{\psi}_{2j+1}(m)
\psi_{2j}(m)+
t_{41}\bar{\psi}_{2j+1}(m+1)\psi_{2j}(m+1)+\right.\nonumber\\
&+&\left.t_{32}\bar{\psi}_{2j+2}(m+2)\psi_{2j+1}(m+2)+
t_{14}\bar{\psi}_{2j+1}(m+1)\psi_{2j+2}(m+1) +\right.\nonumber\\
&+&\left.(1+M_1)\bar{\psi}_{2j}(m)\psi_{2j}(m+1) + 
(1+M_2)\bar{\psi}_{2j}(m+1)\psi_{2j}(m+2) +\right.\nonumber\\
&+&\left.(1-M_{1}^{'})\bar{\psi}_{2j+1}(m)\psi_{2j+1}(m+1) + 
(1-M_{2}^{'})\bar{\psi}_{2j+1}(m+1)\psi_{2j+1}(m+2)\right] +\nonumber\\
&+&\sum_{j,m}\left[\bar{\psi}_{2j+1}(m)\psi_{2j+1}(m) +
\bar{\psi}_{2j}(m)\psi_{2j}(m)\right],
\end{eqnarray}
which is exactly coinciding with the expression (\ref{ACT}) for
zero external magnetic field, if parameters $t_{ij}$ 
relates to $M_{1,2}$ and $M_{1,2}^{'}$ as follows
\begin{eqnarray}
\label{M}
t_{43}=1+M_1;\,\,\,\,\,\,\,\,\,t_{34}=1+M_2,\nonumber\\
t_{21}=1-M_{1}^{'};\,\,\,\,\,\,\,\,\,t_{12}=1-M_{2}^{'}.
\end{eqnarray}

Our aim is the investigation of the Transfer Matrix $T$ in the limit
$t\rightarrow \infty$ after the rescaling of partition function
by $t^{-V}$. This is necessary for 3DIM sign factor.

For  construction of the continuum limit of the Hamiltonian $H$ of the
model defined by
\begin{equation}
\label{T}
T=e^{-H}=:e^{-H_{B_1}}::e^{-H_{B_2}}:,
\end{equation}
we need to analyze his spectrum. Let  make 
 Furies transformation
of the fermionic fields $c_{2i}$ and $c_{2i+1}$ in (\ref{T})
separately
\begin{eqnarray}
\label{FU}
c_{2j+1}&=&\frac{1}{\sqrt{L}}\sum_p c_{1,p} e^{i(2j+1)p}\nonumber\\
c_{2j}  &=&\frac{1}{\sqrt{L}}\sum_p c_{2,p} e^{i 2j p}
\end{eqnarray}
where $p=\frac{2\pi}{L}l,\,\, l= 0,....L/2-1$ are the momentums,
which
corresponds to  Bloch states in the problem
with the translational invariance on two lattice spacing. We have two 
bands in the reduced Brillouin zone. It is obvious that the full
Transfer Matrix $T$ can be represented as
\begin{equation}
\label{TP1}
T=\prod_p:T_{1 p}:: T_{2 p}: =\prod_p T_p ,
\end{equation}
where $T_{1 p}$ and $T_{2 p}$ are the Furies transforms of
$e^{-H_{B_1}}$ and $e^{-H_{B_2}}$ correspondingly.

Our aim is now to represent the product of normal ordered
forms of $T_{1 p}$ and $T_{2 p}$ in the expression (\ref{TP1})
for $T_p$ as
\begin{equation}
\label{TP}
T_p={\cal D} e^{\vec{\varepsilon}_p \vec{S}+\mu(n_{1,p}+n_{2,p})}=
{\cal D} e^{-H_p}
\end{equation}

Here the operators
\begin{equation}
\label{UU}
S^+=c^+_{2,p}c_{1,p};\,\,\,\, S^-=c^+_{1,p}c_{2,p};\,\,\,\,
 S^3=c^+_{2,p}c_{2,p}-c^+_{1,p}c_{1,p}
\end{equation}
form an $Sl_2$ algebra and can be represented as Pauli matrices,
while
\begin{equation}
\label{NN}
n_{i,p}=c^+_{i,p}c_{i,p}, \,\,\, \i=1,2
\end{equation}
are particle number operators.

After some obvious, but not trivial calculations we
will have for $\varepsilon_p = ({\vec\varepsilon}^2_p)^{1/2}$
the chemical potential $\mu$ and ${\cal D}$ following expressions
\begin{eqnarray}
\label{E1}
{\cal D}&=& 2 t_{12}t_{21}   \nonumber\\
e^{2\mu}&=&\frac{t_{34}t_{43}}{t_{12}t_{21}}\nonumber\\
\cosh{\varepsilon_p}&=&\frac{1+\Delta_1 \Delta_2 +
t_{23}t_{32} e^{2ip} +t_{14}t_{41} e^{-2ip}}
{2(t_{12}t_{21}t_{34}t_{43})^{1/2}}.
\end{eqnarray}
where
\begin{eqnarray}
\label{E16}
\Delta_1=t_{21}t_{43}-t_{41}t_{23},\nonumber\\
\Delta_2=t_{12}t_{34}-t_{14}t_{32}.
\end{eqnarray}

In the Appendix we will represent an alternative way of obtaining
of this formulas.

Diagonalization of the Hamiltonian $H_p$ in (\ref{TP}) is easy 
to realize considering it as an $Sl_2$ algebra element. It
is clear that
\begin{equation}
\label{HA}
H_d=-\varepsilon_p S^3 - \mu(n_{1,p}+n_{2,p}),
\end{equation}
hence the dispersion relations for two bands of the
spectrum are
\begin{eqnarray}
\label{E2}
E_{1,p}&=&(\varepsilon_p-\mu),\nonumber\\
E_{2,p}&=&(-\varepsilon_p-\mu).
\end{eqnarray}

Though we start chapter by talking about absence of the external
$U(1)$ field but it is easy to follow that our
expressions remain true for arbitrary double lattice spacing
periodic external fields. In this case simply the hopping
parameters $t_{ij}$ are complex and the phases represent the
$U(1)$ field. In the gauge used in (\ref{KIKO}) we will have the 
same expression  (\ref{E1}) for the spectrum, but where now
\begin{eqnarray}
\label{EN1}
\Delta_1=\omega^{-2}t_{21}t_{43}-\omega^{2}t_{41}t_{23},\nonumber\\
\Delta_2=\omega^{2}t_{12}t_{34}-\omega^{-2}t_{14}t_{32}.
\end{eqnarray}

The expression (\ref{XC}) for the critical line follows from the
(\ref{E1}) and (\ref{EN1}) as a condition $\varepsilon_p =0$.

Up to now we were considering the spectrum  in general 
case, but for our loop gas model in a fully packed regime
it is necessary to consider homogeneous
$t_{ij} \rightarrow t \rightarrow \infty$ limit. As immediately
follows from the (\ref{E1}), this limit is singular and needs
to be analyzed properly. 

Let as go to the limit $t \rightarrow \infty$ together with
$\Phi \rightarrow 0$ always staying on the critical line (\ref{XC}),
that is
\begin{eqnarray}
\label{E19}
t_{23}=t_{32}=t_{14}=t_{41}=
t_{21}=t_{12}=t_{34}=t_{43}=t={1 \over 2\sin{\Phi \over 4}}.
\end{eqnarray}
Then, it easy to see from  (\ref{E1}), that in the limit
$\Phi \rightarrow 0$ we will have $\mu=0$ and
\begin{eqnarray}
\label{E22} 
\cosh{\varepsilon_p} = 2+ \cos 2p.
\end{eqnarray}

In the next section we will argue why this particular scaling
limit was chosen.

From the (\ref{E2}) one can see that the spectrum of the
model is gap-less and at $p= \pm {\pi \over 2}$  two bands 
touch each other
with $\varepsilon_{\pm {\pi \over 2}}=0$. Therefore, when 
the ground state is filled
up to Fermi level corresponding to  $\pm \pi/2$, the spectrum 
is linear near that point
\begin{equation}
\label{L1}
\varepsilon_p = \pm 2 k,
\end{equation}
and we can construct $SO(2)$ invariant field theory. This 
corresponds to half filling of the states. The situation is similar 
to one in the Luttinger liquid \cite{MAH}, but it is necessary
to calculate the central charge of this massless excitations.

The spectrum was analyzed in \cite{DD} in connection
with Hamiltonian walk problem by calculation of the partition 
function and there was established, that it corresponds
to $(0,1)$ spin fermionic system with central charge   
$c= -2$. We will confirm now this result by calculating
of the finite size corrections to ground state energy
of the model \cite{CARD}.

Following  \cite{DD} we use the Euler-MacLaurin formula for
the approximations of finite sums
\begin{eqnarray}
\label{EM}  
E_0&=& \sum_p \varepsilon_p= L/{2\pi}\int^{\pi}_{0}\varepsilon_pdp+
\nonumber\\
&+&\frac{2\pi}{12L}\left(\varepsilon'(\pi/2)-\varepsilon'(-\pi/2)\right)+...,
\end{eqnarray}
where $\varepsilon_p$ is the negative energy branch of the spectrum
and $\varepsilon'(\pi/2)$ and $\varepsilon'(-\pi/2)$ are the velocities of 
right(left)
excitations at the Fermi level, equal to
\begin{equation}
\label{VL}
\varepsilon'(\pi/2)=-\varepsilon'(-\pi/2)= 2.
\end{equation}

Comparing (\ref{EM}) with the finite size correction formula
of excitations with central charge $c$ 
\begin{equation}
\label{C1}
E=-L\varepsilon_0-\frac{\pi v}{6L}c,
\end{equation}
one will obtain $c=-2$ and hence, we have a fermionic system of spins 
(0,1) \cite{FSM}.

It is known, that (0,1) spin system is just the Faddeev-Popov
ghost fields for area preserving transformations.

As it was stressed before, for the partition function of the sign-factors
model (let us mention here again that for the full sign-factor we should
add to model additional vertex operators as perturbations, which 
corresponds to Whitney singularities) we should 
absorb $t$ into $\psi, \bar\psi$  by the scale transformations
\begin{equation}
\label{E24}
\psi^{'}= t^{1/2}\psi,\quad  \bar\psi^{'}= t^{1/2}\bar\psi
\end{equation}
and appropriately rescale the partition function 
$Z(\Phi \rightarrow 0)$ by
the $t^{L N}$ due to fermionic measure in
(\ref{ZET}), where $L N$ is the
number of fermionic degrees of freedom.Then we will come to
\begin{equation}
\label{E25}
{1 \over t^{L N}} Z(\Phi \rightarrow 0)={{\cal D}^{{LN \over 4}} \over 
t^{L N}} Z(c=-2) = (4 \sin{\Phi \over 4})^{{LN \over 4}} Z(c=-2).
\end{equation}

Analyzing the $Det{\cal H}(\vec{k})$ in a fully packed phase
$m \rightarrow 0$ one can easily see, that 
$(4 \sin{\Phi \over 4})^{{LN \over 4}}$ behavior
(which $\rightarrow 0$ with $\Phi \rightarrow 0$) appeared
in the answer as a result of phase factor interference in the
sum over all possible coverings of the surface by loops, while
 any particular representative of this sum  is finit in
the limit $\Phi \rightarrow 0$. The ratio of $Det{\cal H}(\vec{k})$
over $(4 \sin{\Phi \over 4})^{{LN \over 4}}$, which is $Z(c=-2)$,
represents in the limit of the fully packed phase 
the sum over the variety of coverings of space by curves
 with the fixed number of loops.

It is worth to mention in the end of this section, 
that in case of following choose 
of parameters
\begin{eqnarray}
\label{CCP}
t_{41}=t_{12}=t_{23}=-t_{34}= \tanh{\theta},\nonumber\\
t_{14}=t_{21}=t_{32}=-t_{43}= {1 \over \cosh{\theta}},
\end{eqnarray}
the transfer matrix constructed here in (\ref{TP1}-\ref{E16})
will be field theory realization of Chalker and Coddingtons
 transfer matrix for the  scattering  edge excitations in the 
Hall effect. They again form $(0,1)$-spin system with the
central charge $c=-2$ \cite{AS}.

%%%%%%%%%%%%%%%%%%%%%%%%%-------------------------------------------
%%%%%%%%%%%%%%%%%%%%%-----------------------------------------------
%%%%%%%%%%%%%%%%%%%5------------------------------------------------

\section{Interaction with the External Magnetic Field}
\indent

In this section we will reproduce external $U(1)$ field and
background fluxes via two scalar fields.

First, let us introduce dual Manhattan Lattice $\widetilde{ML}$
with sites in the middle points of the $ML$ plaquettes
(A and B points in Fig.1) and
arrow structure with vectors $\vec{\nu}$ on the links.
Vectors $\vec{\mu}$ and $\vec{\nu}$ on $ML$ and $\widetilde{ML}$
are dual each other and intersects perpendicularly in
the middle points of the corresponding links.

Consider scalar fields $\varphi(\vec{n})$ and 
$\bar{\varphi}(\vec{n'})$ ($\vec{n}\in ML, 
\vec{n'}\in \widetilde{ML}$)
on the $ML$ and $\widetilde{ML}$ correspondingly and define
following action for them
\begin{equation}
\label{E13}
S(\varphi,\bar{\varphi})={1 \over \pi}
\sum_{\vec{n},\vec{n'} \atop \vec{\mu}, 
\vec{\nu}} \left(\varphi(\vec{n}+\vec{\mu})-\varphi(\vec{n})\right)
\left(\bar{\varphi}(\vec{n'}+\vec{\nu})-
\bar{\varphi}(\vec{n'})\right).
\end{equation}

In the (\ref{E13}) $\vec{n}$ and $\vec{n'}$ are neighbor sites
on $ML$ and $\widetilde{ML}$ correspondingly, vectors $\vec{\mu}$
and $\vec{\nu}$  exits from $\vec{n}$ and $\vec{n'}$ and crosses
each other in the middle points of the links.

By introducing $\varphi_1(\vec{n})$ and $\varphi_2(\vec{n})$ as
\begin{equation}
\label{E14}
\varphi_1(\vec{n})={1 \over 2}\left(\varphi(\vec{n})+
\bar{\varphi}(\vec{n})\right),\quad \bar{\varphi}_2(\vec{n})
={1 \over 2}\left(\varphi(\vec{n})-
\bar{\varphi}(\vec{n})\right)
\end{equation}
we can separate the holomorphic $\varphi_{1,2}(z), (z=n+im)$
and antiholomorphic $\varphi_{1,2}(\bar{z}), (\bar{z}=n-im)$
parts in each of the fields, which will have following 
correlators \cite{SL}
\begin{eqnarray}
\label{COR1}
\langle\varphi_1(z)\varphi_1(w)\rangle&=&-G(\left|z-w\right|)- iarg(z-w),
\nonumber\\
\langle\varphi_2(z)\varphi_2(w)\rangle&=&+G(\left|z-w\right|)+ iarg(z-w)
\end{eqnarray}
In the continuum limit correlators transforming into ordinary correlators
of free scalar fields in 2D.
\begin{equation}
\label{COR3}
G(\left|z-w\right|)_{\epsilon \rightarrow 0} \rightarrow 
\ln\left|z-w\right|,
\end{equation}
and hence
\begin{equation}
\label{COR2}
G(z-w)_{\epsilon \rightarrow 0} \rightarrow \ln(z-w).
\end{equation}

Let us introduce now following vertex operators
\begin{eqnarray}
\label{VO}
V_A(z,\bar{z})&=&e^{i\frac{Q_1}{4\pi}\Omega \varphi_1+
i\frac{Q_2}{4\pi}\Omega \varphi_2},\nonumber\\
V_B\left(z,\bar{z}\right)&=&e^{-i
\frac{Q_1}{4\pi}\Omega \varphi_1-
i\frac{Q_2}{4\pi}\Omega \varphi_2},
\end{eqnarray}

The charges $Q_{1,2}$ in (\ref{VO}) have to be
defined whereas $\Omega(w)=R(w)\epsilon^2$ ($\epsilon^2$ is the
lattice spacing) can be
interpreted as solid angle of the plaquette, induced by background
curvature $R_b(w)$.
We define the following background curvature in this case
\begin{equation}
\label{DR}
\Omega_0 (w)=\cases{+2\pi \quad\quad w\in A \cr
-2\pi \quad\quad w\in B\cr}
\end{equation}

Let us define now a new fermionic fields in the sites $z$ of the
original $ML$ lattice
\begin{eqnarray}
\label{FF}
\psi_1(z)&=&e^{i\frac{Q_1}{2}\varphi_1(z)+
i\frac{Q_2}{2}\varphi_2 (\bar{z})}C_1(z,\bar{z}) = V_1(z,\bar{z})
C_1(z,\bar{z}),\nonumber\\
\psi_2(z)&=&e^{-i\frac{Q_1}{2}\varphi_1(z)-i\frac{Q_2}{2}\varphi_2
(\bar{z})} C_2(z,\bar{z}) = V_2(z,\bar{z})C_2(z,\bar{z}),\\
\psi_3(z)&=&e^{i\frac{Q_1}{2}\varphi_1(z)+
i\frac{Q_2}{2}\varphi_2(\bar{z})}C_3(z,\bar{z}) = V_3(z,\bar{z})C_3(z,\bar{z}),
\nonumber\\
\psi_4(z)&=&e^{i\frac{Q_1}{2}\varphi_1(z)-
i\frac{Q_2}{2}\varphi_2(\bar{z})}C_4(z,\bar{z}) = V_4(z,\bar{z})C_4(z,\bar{z}).
\nonumber
\end{eqnarray}

In the expression (\ref{FF}) $\varphi_1(z)$ and
$\varphi_2(\bar{z})$ are the holomorphic and antiholomorphic
parts of the scalar fields $\varphi_1$ and $\varphi_2$
correspondingly.
 The $\bar{\psi}_i(z)$ fields are defined by complex conjugation.
The necessity of the introduction of the two $\varphi_1$ and
$\varphi_2$ fields is dictated by the number of degrees of freedom
of original $C_{i}(z)$ fermionic fields.
Also, as we will see later, the presence of two scalar fields
$\varphi _{1,2}$ with opposite sign of kinetic energy and
appropriate choose of $Q_{1,2}$, insures
the cancellation of total central charge of the theory, which we
need in the model (\ref{ACT}, \ref{STS1}) by definition. The choice of
coefficients $\frac{Q_{1,2}}{2}$ in the exponents of (\ref{FF})
is dictated by
coincidence of parameters of the Jacobian of
$C_i\rightarrow \psi_i$ transformations
with the background charges  $\frac{Q_{1,2}}{4\pi }\Omega $
in $V_A, V_B$ (see (\ref{VO})).

We would like to show, that the partition function (\ref{STS1}) is
equal to
\begin{equation}
\label{SS}
Z(0)=\int \prod_{i=1}^{4} \prod_{\vec{n}}\,d\bar{C}^{i}_{\vec{n}} dC_{\vec{n}}^i
\prod_{z\in {\rm sites}\atop\alpha =1,2}d\varphi _\alpha  (z)
d \varphi _\alpha
(\bar{z}) \prod_{z\in{\rm plaquettes}}V_A(z,\bar{z}) V_B(z,\bar{z})
e^{S(\varphi_1,\varphi_2)+A(\psi_i;0)}
\end{equation}
where $ S(\varphi_1\varphi_2)$ is the action of the free scalar
fields $\varphi_1,\varphi_2$ with correlators (\ref{COR1})
and  $A(\psi_i;0)$ is the
fermionic action (\ref{E12}) on $ML$, but with fields (\ref{FF}), placed in
the zero external magnetic field.

Let's first make integration over $C^i_{\vec{n}}$. Because of
Grassmann character of variables $C^i_{\vec{n}}$, the nonzero
contribution in integral (\ref{SS}) can have only terms, where at each
site $z$ of $ML$ we have one pair $\bar{C}^i_{\vec{n}} C^i_{\vec{n}}$.
The expansion of $e^{A(\psi_i;0)}$ in  (\ref{SS}) over $\psi_i$
produces two type of nonzero products of $\bar{\psi}_i(z) \psi_i(z)$
at the each site of ML. The first come as a product of hopping terms
in $A(\psi_i;0)$ along closed oriented contours $\gamma_{\sigma},
(\sigma =1,...,k)$ on ML. The second come as a product of mass terms
$M \bar{\psi}_i(z) \psi_i(z)$ in $A(\psi_i;0)$ along another closed
oriented contours $\bar{\gamma}_\tau, (\tau = 1,...,l)$, which
covers the complement to $\gamma_{\sigma}$ sites. The sum of contours
$\gamma_{\sigma} U  \bar{\gamma}_{\tau}$ densely covers ML by the
all possible ways. Hence the term, which have nonzero contribution
into partition function $Z(0)$ is the following
\begin{eqnarray}
\label{BN}
{\cal P}=\sum_{{\rm all\,\,possible\atop
coverings}}\prod^{k}_{\sigma =1}\prod_{\vec{n}\in \gamma_\sigma} 
\rho e^{\Gamma_{\vec{n},\vec{\mu}}}
\prod_{z\in {\gamma}_\sigma }
\bar{\psi}_i(z)\psi_j(z+\mu)\prod^{l}_{\tau =1}
\prod_{z\in \bar{\gamma}_\tau }\bar{\psi}_i(z)\psi_i(z)=
\nonumber\\
\sum_{{\rm all\,\,possible\atop
coverings}}\prod^{k}_{\sigma =1} \prod_{\vec{n}\in \gamma_\sigma} \rho
e^{\Gamma_{\vec{n},\vec{\mu}}} \prod_{z \in \gamma_\sigma }
\bar{C}_i(z)C_j(z+\mu)V_i^+(z)V_j(z+\mu)
\prod^{l}_{\tau =1} \prod_{z\in \bar{\gamma}_\tau }
\bar{C}_i(z) C_i(z)
\end{eqnarray}

The integration over $\varphi_{\alpha}$ in (\ref{SS}) means the
calculation of average of the product of operators $V_A,V_B$
and $\varphi _{\alpha}$ dependent  parts of the fields $\psi_{i}(z)$ in
(\ref{BN}) (see definition of $\varphi_i$ in (\ref{FF})).

Therefore
\begin{eqnarray}
\label{BNN}
Z(0)&=&\int \prod_{i=1}^{4} \prod_{\vec{n}}\,dC_{\vec{n}}^i\,
d \bar{C}_{\vec{n}}^i\,
\tilde{Z},\nonumber\\
\tilde{Z}&=&\langle\prod_{z \in\ {\rm plaquettes}}V_A(z,\bar{z})
V_B(z,\bar{z}){\cal P}\rangle_{\varphi_1\varphi_2}Z_{\varphi} 
=Z_1Z_2Z_3Z_{\varphi}
\end{eqnarray}
where $Z_{\varphi}$ is the partition function of the free
scalar fields $\varphi_{1,2}$
\begin{eqnarray}
\label{E26}
Z_{\varphi}=\int {\cal D}\varphi_1 {\cal D}\varphi_2 e^{
-S(\varphi_1,\varphi_2)}
\end{eqnarray}

The variable $Z_1$ in (\ref{BNN}) consists only of mutual correlators of
vertex operators $V_{A,B}$ and equal
\begin{eqnarray}
\label{Z1}
Z_1=\exp\left(\sum_{z,w\in\ {\rm plaquettes} \atop {\rm of} \,  ML }
\left(\left(\frac{Q_1}{4\pi}\right)^2-
\left(\frac{Q_2}{4\pi}\right)^2\right)
        \Omega(z)G(|z-w|) \Omega(w)\right)
\end{eqnarray}

The $Z_2$ consists of mutual correlators of $\varphi_{1,2}$,
dependent parts of Grassmann fields $\psi_i$
in (\ref{Z1}) and (\ref{BNN}).
\begin{eqnarray}
\label{JK}
&&Z_2=\exp\left(\sum_{z,w\in\ {\rm sites} \atop {\rm of} \,  ML }
\left(\left(\frac{Q_1}{2}\right)^2-
\left(\frac{Q_2}{2}\right)^2\right)
        M(z)G(|z-w|) M(w)\right)
\end{eqnarray}
where
\begin{equation}
\label{KL}
M(z)=\cases{+1,\quad\quad{\rm for}\quad \quad  \psi_{1,3} ;\cr
-1,\quad\quad {\rm for} \quad \quad \psi_{2,4},\cr}
\end{equation}
We see from (\ref{Z1},\ref{JK}) that condition $Q_1=Q_2=Q$ is insuring
the full cancellation of $G(|z-w|)$ dependent parts in $Z_{1,2}$, giving
\begin{equation}
\label{LP}
Z_1=Z_2=1
\end{equation}
The term $Z_3$ in (\ref{BNN}) is defined by the correlators of
$V_{A,B}$  with $\varphi _{\alpha}$ dependent parts of $\psi_i$.
It is
easy to see, that
\begin{eqnarray}
\label{UI}
Z_3&=&\sum_{{\rm all\,\,possible\atop coverings}}\prod_{\sigma=1}^{k}
\prod_{\vec{n}\in \gamma_\sigma} \rho e^{\Gamma_{\vec{n},\vec{\mu}}}
\prod_{z\in {\gamma}_\sigma}\bar{C}_i(z) K^+(z)
K(z+\mu) C_j(z+\mu),
\end{eqnarray}
where
\begin{equation}
\label{DK}
K(z)=e^{\frac{Q}{4\pi }\frac{Q}{2}\sum_{w}\Omega (w)\left[G(z-w)-
G(\bar{z}-\bar{w})\right]}.
\end{equation}
and the sum is taken over middle points of all (A and B) plaquettes.

The product $K^{+}(z)K(z+\mu)$ in
(\ref{UI}), being element of $U(1)$ group, is nothing but $U(1)$
gauge field part in the action (\ref{ACT}) in fixed gauge.
The integration over fermionic fields $C_i(z), \bar{C}_i(z)$ in partition
function is simple and produces (-1) for the each closed contour
$\Gamma_\sigma$ in the (\ref{UI}) together with the product of expressions
\begin{equation}
\label{KP}
K^{+}(z)K(z+\mu)=e^{\sum_{w}\frac{Q^2}{4\pi}
\Theta_{z,z+\epsilon}(w)\Omega (w)},
\end{equation}
along them. In (\ref{KP}) $\Theta_{z,z+\mu}$ is the looking angle of
the link $(z,z+\mu)$ from the middle point $w$ of some plaquette in ML
(see Fig.2) and is equal to $2\pi$. Finally, by use of (\ref{DR}), 
we have the following expression for  each contour
in (\ref{UI})
\begin{equation}
\label{WA}
\prod_{{\rm along\,c}}K^{+}(z)K(z+\mu)=e^{i\pi Q^2}.
\end{equation}

\begin{center}
\setlength{\unitlength}{1cm}
\begin{picture}(15,8)
%\put(0,1){\vector(1,0){12.5}}
%\put(1,0){\vector(0,1){6}}
\multiput(1,1)(3,0){4}{\circle*{0.2}}
\multiput(1,3)(3,0){4}{\circle*{0.2}}
\multiput(1,5)(3,0){4}{\circle*{0.2}}
\multiput(1,7)(3,0){4}{\circle*{0.2}}
\multiput(5.5,2)(0.3,1){5}{\line(1,3){0.2}}
\multiput(5.5,2)(-0.3,1){5}{\line(-1,3){0.2}}
%\multiput(1.4,1.4)(-0.3,0.6){10}{\line(6,4){0.2}}
%\put(1,1){\line(6,4){6}}
\put(5.5,3.1){\oval(0.7,0.5)[t]}
%\put(2.1,1.3){\vector(0,-1){.3}}
%\put(4.4,4){\oval(1,1)[tr]}
%\put(4.9,4){\vector(0,-1){.5}}
\put(4,7){\vector(1,0){2.9}}
\put(1,3){\vector(1,0){2.9}}
\put(4,3){\vector(0,1){1.9}}
\put(4,5){\vector(0,1){1.9}}
\put(0,3){\vector(1,0){0.9}}
\put(7,7){\vector(1,0){2.9}}
\put(10,7){\vector(1,0){0.9}}
%\put(2.4,1.3){\shortstack{$\Theta_z $}}
\put(4,7.3){\shortstack{z}}
\put(5,3.8){\shortstack{$\Theta_{z,z+\epsilon}$}}
\put(7,7.3){\shortstack{$z+\epsilon$}}
%\put(0.5,.5){\shortstack{$O$}}
\put(6.5,0){\shortstack{Fig.2.}}
\end{picture}
\end{center}

In order to have a phase factor $e^{2\pi i p/q}$ which we need,
one should take
\begin{equation}
\label{QQ}
Q^2=2p/q
\end{equation}

One can obtain (\ref{WA}), taking into account (\ref{DR}) and the
fact, that only points inside the contour $C$ in
(\ref{KP}) contribute to the phase and due to the Manhattan
nature of the lattice we always have an extra $+$ or $-$ value for
the $\Omega(w)$. An additional $-$ sign for each closed contours 
appeared because of the fermionic measure for the loops.

Therefore, after integration over $\varphi_{\alpha}$,
the whole partition function $Z(0)$ reduces to the product of 
factors
\begin{equation}
\label{ZZ}
Z(0)=Z_{\varphi}\int \prod_{i=1}^{4} \prod_{\vec{n}}\,dC_{\vec{n}}^i\,
d \bar{C}_{\vec{n}}^i \prod_{{\rm all\,\, lattice\,\atop points}}
\bar{C}_i(z) C_i(z) \sum_{{\rm all\,\,coverings}}
\prod_{\sigma=1}^{k}(-e^{i\Phi_\sigma})
\prod_{\vec{n}\in \gamma_\sigma}\rho e^{\Gamma_{\vec{n},\vec{\mu}}}.
\end{equation}

In the limit ${m \over t} \rightarrow 0$ the last component of the
product in (\ref{ZZ}) will be cancelled by the rescaling factor
$t^{-LN}$. The $C, \bar C$-part and the last term of this expression 
represents
the variety of  coverings of the lattice by fermionic curves
 with the fixed number of loops
and, as it was pointed out in the section 2, is equal to $Z(c=-2)$ 
in the limit of fully packed phase ${m \over t} \rightarrow 0$.
 As for partition function of free
scalars $Z_{\varphi}$ in continuum limit, in 
$SO(2)$ rotation invariant regularization we have
\begin{equation}
\label{E27}
Z_{\varphi} =\exp\left(-{1 \over 48 \pi^2} 
\int\Omega(z)G(|z-w|)\Omega(w)\right) 
\end{equation} 
for each of them,
which corresponds to central charge $c=1$. Therefore,
the fermionic and bosonic parts the expression (\ref{ZZ})
cancel each other and we are left with
\begin{equation}
\label{ZZ1}
Z(0)=\sum_{{\rm all\,\,coverings}}
\prod_{\sigma=1}^{k}(-e^{i\Phi_\sigma}),
\end{equation}
where $\Phi_\sigma$ is the $U(1)$ flux inside the contour
$\gamma_\sigma$ and this expression indeed coincides
with the (\ref{STS1}) in the ${m \over t} \rightarrow 0$ limit.
 Now it is clear why in the section 3 we
have considered a particular scaling limit. Only in that 
unique limit we can have cancellation of bosonic and fermionic
anomalous parts and our partition function will represent the 
fully packed limit of the model, defined  in (\ref{STS1}).

Expression (\ref{E27}) together with (\ref{Z1}) shows that the
central charges of scalars are $1 \pm 3 Q^2$. 

Finally let us make the following remark. Introduced
 two free scalars with  the opposite statistics 
are equivalent
to two dimensional vector field
\begin{equation}
\label{E18}
A_{\alpha}=\partial_{\alpha}\varphi_1 + 
\epsilon_{\alpha \beta}\partial_{\beta}\varphi_2
\end{equation}
with the statistical weights equal 
\begin{equation}
\label{E17}
P({A_{\alpha}})= \exp\left(\int A_{\alpha}^2\right) .
\end{equation}

%------------------------------------------------
%---------------------------------------------
%------------------------------------------------
\section{Continuum Limit of the Model}
\indent
In the previous sections
we have represented model as a free fermionic spin (0.1) system
with central charge $c=-2$, interaction of 
which with the external magnetic field can be constructed via two
scalars of opposite statistics and vacuum charge $Q^2=2p/q$. 
%----------------------------------------------
For the spin (0,1) fermionic system the action is 
\begin{equation}
\label{DA}
A(\psi_i, 0) = {1 \over \pi}\int\left(\bar\psi_L\partial\psi_L +
\bar\psi_R\bar\partial\psi_R\right) 
\end{equation}
where $\psi_R(\psi_L)$ corresponds to $\psi_{2,4}(\psi_{1,3})$
fermions.

The formulation of continuum limit of the vacuum charge parts 
of the scalar fields
in the action, which corresponds to background flux operator,
is straightforward (see expressions of vertex
operators in (\ref{SS}))
\begin{eqnarray}
\label{VCP}
\prod_{z\in{\rm plaquettes}}V_A(z,\bar{z}) V_B(z,\bar{z})&=&
\exp\left(\sum_{w}i\frac{Q}{4\pi}\Omega (w)
(\varphi_1+\varphi_2)\right)=\nonumber\\
&=&\exp\left(i\frac{Q}{4\pi}\int\,d^2
w R(w)(\varphi_1+\varphi_2)\right)
\end{eqnarray}

One can generalize the model (4) for the random surfaces. 
All closed random $ML$-s are in one to one correspondence
with the closed surfaces in regular $3D$ lattice.
Following \cite{KS} they can be constructed as follows.
Let us consider middle points of the links on such surface in
regular lattice as sites of the new lattice which we would
like to construct and connect 
them with links. Now, by drawing arrows on these links
as in Fig.3 we will complete the construction. 
The $B_1$ and $B_2$ plaquettes in Fig.3 coincides after 
rotation on $\pi/2$. 

\begin{center}
\setlength{\unitlength}{.8cm}
\begin{picture}(10,6)
\put(1,1){\line(0,1){4}}
\put(5,1){\line(0,1){4}}
\put(9,1){\line(0,1){4}}
\put(1,1){\line(1,0){4}}
\put(5,1){\line(1,0){4}}
\put(1,5){\line(1,0){4}}
\put(5,5){\line(1,0){4}}
\put(3,5){\vector(-1,-1){1}}
\put(2,4){\line(-1,-1){1}}
\put(3,5){\vector(1,-1){1}}
\put(4,4){\line(1,-1){1}}
\put(3,1){\vector(1,1){1}}
\put(4,2){\line(1,1){1}}
\put(3,1){\vector(-1,1){1}}
\put(2,2){\line(-1,1){1}}

\put(5,3){\vector(1,1){1}}
\put(6,4){\line(1,1){1}}
\put(7,5){\line(1,-1){1}}
\put(9,3){\vector(-1,1){1}}
\put(9,3){\vector(-1,-1){1}}
\put(8,2){\line(-1,-1){1}}
\put(6,2){\line(1,-1){1}}
\put(5,3){\vector(1,-1){1}}
\put(3,3){\shortstack{\circle*{0.1}}}
\put(3.3,3){\shortstack{$B_1$}}
\put(7,3){\shortstack{\circle*{0.1}}}
\put(7.3,3){\shortstack{$B_2$}}
\put(4.5,0){\shortstack{Fig.3.}}
\end{picture}
\end{center}

The $B_{1,2}$ faces of $ML$ are always quadrangles, while
$A_{1,2}$ type faces, which formed around sites of
the original $3D$ regular lattice, can be arbitrary 
$n$-angle, with $n$= 3,4...9 (in case of regular target
lattice). Hence, the  curvature of  $A_{1,2}$
faces can be changed and become
\begin{equation}
\label{E29}
\Omega_n=\frac{\pi }{2}(4-n)
\end{equation}

We would like to have a $U(1)$ flux in the
$n$-angle face on random $ML$, equal to $\frac{\pi }{2}n$, which is
proportional to area of the $n$-angle. Then
the corresponding solid angle can be written as
\begin{equation}
\Omega (w)=\cases{\Omega_0-\Omega_n(w)=2\pi-\Omega_n(w) 
\quad\quad w\in A;\cr
-\Omega_0=-2\pi \quad\quad\quad\quad\quad\quad \quad 
\quad\quad  w\in B,\cr}
\end{equation}
and the formula for the loop gas representation of the models
partition function (\ref{STS1}) for random surface will be
modified as follows
\begin{equation}
\label{RAN}
Z=\sum_{{\rm all\,\,possible\atop coverings}} \prod_{\sigma,\tau}
(-\eta(\gamma_\sigma))(-\eta(\bar{\gamma}_\tau))
\Phi(\gamma_\sigma)\Phi(\bar{\gamma}_\tau),
\end{equation}
where 
\begin{equation}
\label{RAO}
\eta(\gamma_\sigma)=exp\left(2 \pi \frac{p}{q} \int_D \Omega\right),
\end{equation}
with $\gamma_\sigma = \partial D$.
So we see, that response of the model on inclusion of gravity
is simply consists in appearance of the gravitational curvature
of the random surface in the vacuum charge term (\ref{VCP}) of the
effective action $S$
\begin{equation}
\label{LK}
R=R_{{\rm background}}+R_{{\rm grav}}
\end{equation}

Finally we get the action of the defined model for the continuum
limit as
\begin{eqnarray}
\label{FACT}
S&=\frac{1}{\pi}&\int\,d^2z\left[\bar{\psi}_L\partial\psi_L+
\bar{\psi}_R\bar{\partial}\psi_R
+ {1 \over 2}\partial\varphi_1\bar{\partial}\varphi_1-
{1 \over 2} \partial\varphi_2\bar{\partial}\varphi_2-\right.\nonumber\\
&-&i\left.\frac{Q}{8}R(  \varphi_1+\varphi_2)\right].
\end{eqnarray}
Here the fermionic field $\psi$ is the (0,1) spin system with
central charge equal to $-2$, which is necessary to cancel $+2$ part
of the central charge of free fields $\varphi_1$ and $\varphi_2$,
appearing in continuum limit in addition to $3Q^2$, calculated in
(\ref{Z1}).

It is easy to recognize in the obtained action the action for twisted
$N=2$ superconformal theories \cite{MSS,I}, but in order to fix
the concrete type of the theory, the $SU(2)/U(1)$ coupling
constant $k$, it is necessary  also to compare the $U(1)$ current.
The usual $U(1)$ current on lattice is defined by hopping
of fermions and the $z$ (correspondingly $\bar{z}$) component
of such current simply is
\begin{equation}
\label{JA}
J_z= J_{\vec{n},\vec{\mu_1}}+
i J_{\vec{n}+\vec{\mu_1}+\vec{\mu_2},-\vec{\mu_2}} = 
\bar{C}^4_{\vec{n}+\vec{\mu_1}}
(C^1_{\vec{n}}+i C^3_{\vec{n}+\vec{\mu_1}+\vec{\mu_2}}) ,
%\bar{\psi} \psi +i Q \partial\varphi_1
\end{equation}
where $\mu_1$ and $\mu_2$ are the vectors on the $ML$ in $x$ 
and $y$ directions correspondingly. By use of expressions 
(\ref{FF}) for $C^i(z,\bar{z})$, and with the help of 
properties of the vertex operators, it is not hard to find, 
that
\begin{equation}
\label{E28}
J_z=\bar{\psi}_L \psi_L +i Q \partial\varphi_1.
\end{equation} 
This expression of $U(1)$ current, together with the
action (\ref{FACT}), allows to fix a model as twisted \cite{EY}
$N=2$ superconformal theory \cite{MSS, I} at the level
\begin{equation}
\label{E30}
k+2=\frac{2}{Q^2}=\frac{q}{p}
\end{equation}

The connection of N=2 superconformal theories with the $2d$ polymer 
physics, percolation
problem and polymers at the $\Theta$-point was investigated in
\cite{S2,S3}.

On the other side it is known  \cite{MV,OS} that
twisted $N=2$ superconformal theories at the level $k$ are
equivalent to $(p,q)$ minimal models, interacting with $2d$
gravity and are topological. So, the described model of
$\pm 2\pi p/q$ fluxes can be considered as $(p,q)$ minimal
model interacting with $2d$-gravity and as microscopic 
statphysical equivalent for
topological theories. We hope in this way to find a topological
definition for sign factor of $3DIM$.
%%%%%%%%%%%%%%%%%%%%%%%%%%%%%%%%%%
%%%%%%%%%%%%%%%%%%%%%%%%%%%%%%%%%
\section{Acknowledgments}
\indent

I would like to acknowledge A.Polyakov for the collaboration
at the initial stage of the work on this article during my
visits to Princeton University in the falls of 1991 and 1993.

I appreciate very much the discussions with J.Ambjorn, 
H.Babujian, P.van Baal, T.Hakobyan, D.Karakhanyan,
I.Kostov, H.Leutwyler and R.Poghossian.
 I am grateful to Newton Institute of Mathematical Science 
at Cambridge for hospitality, where the part of this work was 
done and Schweizerischer Nationalfond for support in part.                 
%%%%%%%%%%%%%%%%%%%%%%%%%%%%%%------------------------------
%%%%%%%%%%%%%%%%%%%%%%%%%%%%%--------------------------------
%%%%%%%%%%%%%%%%%%%%%%%%%%%%%--------------------------------
\section{Appendix}
\indent

Let us first calculate $Z(0)$, which simply
connected with the determinant of the operator ${\cal H}(\vec{k})$,
defined by eq. (\ref{KIKO}) as
\begin{eqnarray}
\label{A1}
Z(0)&=&\prod_{\vec{k}} det[{\cal H}(\vec{k})] =
\prod_{l,n}\left\{1 + (t_{14}t_{32}-t_{12}t_{34})
(t_{23}t_{41}-t_{43}t_{21})+\right.\nonumber\\
&+& \left.\left(t_{23} t_{32} e^{2i k_x}+ 
t_{14} t_{41} e^{-2i k_x}+t_{43} t_{34} e^{2i k_y}+t_{12} t_{21} 
e^{-2i k_y}\right)\right\},\\
k_x&=&{2\pi \over L}l, \quad l=0,...{L \over 2}-1,\nonumber\\
k_y&=&{2\pi \over N}n, \quad n=0,...{N \over 2}-1.
\end{eqnarray}
 
The aim is to represent $Z(0)$ as
\begin{eqnarray}
\label{A2}
Z(0)= Tr T^{N/2} = \prod_{k_x}Tr(T_{k_x})^{N/2},
\end{eqnarray}
where
\begin{equation}
\label{A3}
T_{k_x}={\cal D} \exp\left((\varepsilon_{k_x}+\mu) n_{1,k_x}+
(-\varepsilon_{k_x}+\mu) n_{2,k_x}\right).
\end{equation}

Then
\begin{eqnarray}
\label{A4}
Z(0)&=&\prod_{k_x} {\cal D}^{N \over 2} \left(
1+ e^{(\mu +\varepsilon_{k_x}){N \over 2}}\right)
\left(1+ e^{(\mu -\varepsilon_{k_x}){N \over 2}}\right)\nonumber\\
&=&\prod_{k_x} 2{\cal D}^{\mu {N \over 2}}\left(
\cosh{{N \over 2}\mu} +\cosh{{N \over 2}\varepsilon_{k_x}}\right).
\end{eqnarray}

Now, if we will use the known trigonometric identity
\begin{eqnarray}
\label{A5}
\cosh{{N \over 2}\varepsilon_{k_x}}+\cosh{{N \over 2}\mu}=
2^{{N \over 2}-1}\prod_{n=0}^{{N \over 2}-1}
\left\{\cosh{\varepsilon_{k_x}}+
\cos (i\mu+{2\pi \over N/2}n)\right\}  
\end{eqnarray}
in the expression (\ref{A4}) for $Z(0)$, and compare it
with the (\ref{A1}), we will find an exact coincidence,
provided that
\begin{eqnarray}
\label{A6}
{\cal D}&=& 2 t_{12}t_{21}   \nonumber\\
e^{2\mu}&=&\frac{t_{34}t_{43}}{t_{12}t_{21}}\nonumber\\
\cosh{\varepsilon_p}&=&\frac{1+(t_{14}t_{32}-t_{12}t_{34})
(t_{23}t_{41}-t_{43}t_{21})}{2(t_{12}t_{21}t_{34}t_{43})^{1/2}}+\\
&+&\frac{t_{23}t_{32} e^{2ip} +t_{14}t_{41} e^{-2ip}}
{2(t_{12}t_{21}t_{34}t_{43})^{1/2}}.\nonumber\\
\end{eqnarray}

In the scaling limit of $Z(0)$, defined by 
eq.(\ref{E19})
and $\Phi \rightarrow 0$, we will have the same
expression for the partition function as for Hamiltonian walk
obtained in \cite{DD}, where was shown, that it coincides with the
partition function of $c= -2$ particles on the torus.

\end{document}